\documentclass{jpsj-suppl}
\usepackage{txfonts} 
\usepackage{graphicx,bm,booktabs}
\usepackage{mathrsfs,amsfonts,amsmath,color}

\title{$\bar{D}\Sigma^*_c$ and $\bar{D}^*\Sigma_c$ interactions and
 LHCb pentaquarks}

\author{Jun He}

\inst{Theoretical Physics Division, Institute of Modern Physics, Chinese Academy of Sciences,
Lanzhou 730000, China\\
Research Center for Hadron and CSR Physics,
Lanzhou University and Institute of Modern Physics of CAS, Lanzhou 730000, China}

\email{junhe@impcas.ac.cn}

\recdate{September 31, 2016}

\abst{Recently, LHCb collaboration reported the observation of two hidden-charmed resonances $P_c(4380)$ and $P_c(4450)$ consistent with hidden-charmed pentaquarks.  We perform a dynamical investigation about the $\bar{D}\Sigma_c^*(2520)$ and $\bar{D}^*\Sigma_c(2455)$ interactions which are described by the meson exchanges in a quasipotential Bethe-Salpeter equation approach. Two poles around $4450$ and $4390$ MeV are produced  from the $\bar{D}^*\Sigma_c(2455)$ interaction with spin parities $3/2^-$ and $5/2^+$, respectively.  The peak for $5/2^+$ state has a comparable hight as that of $3/2^-$ state in the $J/\psi p$ invariant mass spectrum. Another bound state with spin-parity $J^P=3/2^-$ is produced from  the $\bar{D}\Sigma^*_c(2520)$ interaction. Such results suggest that the narrower LHCb pentaquark $P_c(4450)$ can be well interpreted as a  $5/2^+$ $\bar{D}^*\Sigma_c(2455)$ molecular state while the $P_c(4380)$ is a $3/2^-$ $\bar{D}^*\Sigma_c(2455)$ molecular state mixed with other secondary  origins. }

{\kword{Hidden-charmed pentaquark, Quasipotential Bethe-Salpeter equation, Hadronic molecular state}}

\begin{document}
\maketitle

\section{Introduction}

The resonant structure $P_c(4450)$ and $P_c(4380)$ observed in the $J/\psi p$ invariant mass spectrum at LHCb~\cite{Aaij:2015tga} are promising candidates of the pentaquarks which has be expected for a long time in the particle physics community. Before the LHCb experiment, the predictions about the hidden-charmed nucleon resonance have been made in several approaches, such as the molecular state from S-wave interaction of an anticharmed meson and a charmed baryon~\cite{Wu:2010jy,Yang:2011wz}  and the pentaquark in one color singlet~\cite{Yuan:2012wz}. The masses of the LHCb pentaquarks are in the energy region of  a system composed of an anticharmed meson and a charmed baryon considered in Refs.~\cite{Wu:2010jy,Yang:2011wz}. Especially, the two $P_c$ states locate just below the $\bar{D}\Sigma^*_c$ and  $\bar{D}^*\Sigma_c$ thresholds with mass gaps about 5 and 15 MeV, respectively. So it is naturally to assign two LHCb pentaquarks as hadronic molecular states from the $\bar{D}\Sigma^*_c$ and $\bar{D}^*\Sigma_c$ interactions. However, the partial wave analysis at LHCb favors that the $P_c(4380)$ and $P_c(4450)$ carry opposite parities~\cite{Aaij:2015tga}. It makes it very difficult to explain both states as hadronic molecular states from relevant S-wave interactions because $\bar{D}^{(*)}$ and $\Sigma_c^{(*)}$ have opposite and negative parities, respectively. We should try to go beyond S-wave interaction to interpret the LHCb pentaquarks in  molecular state pictuer composed of an anticharmed meson and a charmed baryon.

In the literature, there are only a few studies about  P-wave molecular state~\cite{He:2010zq,Kang:2016zmv}, especially Ref. ~\cite{Kang:2016zmv} where S-wave interaction is forbidden.
In this work, with the help of  a partial wave decomposition on spin parity $J^P$, we will study the molecular states with both positive and negative parities from the $\bar{D}\Sigma^*_c$, $\bar{D}^*\Sigma_c$ and $\bar{D}^*\Sigma^*_c$  interactions in a quasipotential Bethe-Saltpeter equation.  Besides, the effects of the molecular states with different spin parities on the $J/\psi p$ invariant mass spectrum are also discussed.

\section{Bound states from $\bar{D}\Sigma^*_c$, $\bar{D}^*\Sigma_c$ and $\bar{D}^*\Sigma^*_c$  interactions}

We consider the interactions between  an anticharmed meson $D^{(*)}$ and a charmed baryon $\Sigma_c^{(*)}$, which are described by the light meson exchanges including  pseudoscalar ($\pi$ and $\eta$), vector ($\rho$ and $\omega$) and scalar ($\sigma$) meson exchanges. The potential can be obtained with the help of Lagrangians from the heavy quark effective theory~\cite{Cheng:1992xi,Yan:1992gz,Wise:1992hn,Casalbuoni:1996pg}, which are collected in Ref.~\cite{Yang:2011wz}, and the explicit potential can be found in ~\cite{He:2015cea}.
In a study of the $DD^*$ interaction~\cite{He:2015mja}, the $J/\psi$ exchange was included due to the OZI suppression of light meson exchange. In the interactions considered here,  such OZI suppression does not exist. Hence,only  light meson exchanges are included in calculation with an argument that the exchanges of heavy-mass mesons should be suppressed.

Here, we adopt a Bethe-Saltpeter approach with a spectator quasipotential approximation~\cite{Gross:1991pm}, which was explained explicitly in the appendices of Ref.~\cite{He:2015mja} and applied to studied the $\Lambda(1405)$, the $Z_c(4430)$, the $N(1875)$, and the $Z(3900)$~\cite{He:2015mja,He:2014nya,He:2015cca,He:2014nxa,He:2015yva}, to search the possible bound states related to the $P_c$ states. The spectator quasipotential approximation is unitary and covariant~\cite{Gross:1991pm}. Because spin parity $J^P$ instead of orbital angular momentum $L$ are good quantum numbers if the formulism is relativistic~\cite{Chung}, the partial wave decomposition is done into the quantum number $J^P$ directly in this work. It is an advantage of our method because the experiment result is usually provided with spin parity $J^P$. The Bethe-Saltpeter equation for partial-wave amplitude with fixed spin-parity $J^P$ reads ~\cite{He:2015mja},
\begin{eqnarray}
{\cal M}^{J^P}_{\lambda'\lambda}({\rm p}',{\rm p})
&=&{\cal V}^{J^P}_{\lambda',\lambda}({\rm p}',{\rm
p})+\sum_{\lambda''}\int\frac{{\rm
p}''^2d{\rm p}''}{(2\pi)^3}
{\cal V}^{J^P}_{\lambda'\lambda''}({\rm p}',{\rm p}'')
G_0({\rm p}''){\cal M}^{J^P}_{\lambda''\lambda}({\rm p}'',{\rm
p}), \label{Eq: BS_PWA}
\end{eqnarray}
where the sum extends only over nonnegative helicity $\lambda''$.
The partial wave potential is written as
\begin{eqnarray}
{\cal V}_{\lambda'\lambda}^{J^P}({\rm p}',{\rm p})
&=&2\pi\int d\cos\theta
~[d^{J}_{\lambda\lambda'}(\theta)
{\cal V}_{\lambda'\lambda}({\bm p}',{\bm p})+\eta d^{J}_{-\lambda\lambda'}(\theta)
{\cal V}_{\lambda'-\lambda}({\bm p}',{\bm p})],
\end{eqnarray}
where the initial and final momenta are chosen as ${\bm p}=(0,0,{\rm p})$  and ${\bm p}'=({\rm p}'\sin\theta,0,{\rm p}'\cos\theta)$. The $d^J_{\lambda\lambda'}(\theta)$ is the Wigner d-matrix. An exponential
regularization will be adopted in this work by introducing a form factor $F(k^2)=e^{-(k^2-m^2)^2/\Lambda^4}$ to the off-shell particle in the propagator. Besides, to compensate the
off-shell effect of exchange meson, a form factor  in a form of  $f(q^2)=\Lambda^8/(\Lambda^2-q^2)^4$ is introduced.

The partial wave Bethe-Salpeter equation can be solved by transforming the integral equation to a matrix equation. The pole of scattering  amplitude $\cal M$ is located at $z=M+i\Gamma/2$ where $|1-V(z)G(z)|=0$~\cite{He:2015mja}.  In this work, the cutoff $\Lambda$ is scanned from 0.5
to 5 GeV to search for the poles which correspond to the bound states from the $\bar{D}\Sigma^*_c$, $\bar{D}^*\Sigma_c$ and $\bar{D}^*\Sigma^*_c$  interactions, and only states with  half isospin,which are related to the $P_c$ state observed at LHCb, will be considered.
We collect the bound states relevant to the $P_c(4380)$ and the $P_c(4450)$ based on their masses as below,
\begin{flushleft}
\begin{tabular}{llll}
$P_c(4380)$:& $\bar{D}\Sigma_c^* \ [3/2^-, 0.7{\rm-}1.4]$,& $\bar{D}\Sigma_c^*\ [3/2^+, 2.8{\rm -}5.0]$, & $\bar{D}^*\Sigma_c [3/2^-, 3.0{\rm-}3.7]$; \nonumber\\
$P_c(4450)$:& $\bar{D}^*\Sigma_c [5/2^+, 2.7{\rm-}2.8]$,& $\bar{D}^*\Sigma_c [5/2^-, 2.75{\rm -}2.85]$, & $\bar{D}^*\Sigma^*_c [5/2^+, 2{\rm-}2.1]$.  \nonumber
\end{tabular}
\end{flushleft}
The values in the bracket are spin parities of the system and the cutoffs $\Lambda$ in the unit of GeV which produce the experimental mass within uncertainties.

For lower LHCb pentaquark $P_c(4380)$, there are three candidates, two of which are produced from the $\bar{D}\Sigma^*_c$ interaction and one of which from  the $\bar{D}^*\Sigma_c$ interaction. Two of the candidates for higher LHCb pentaquark $P_c(4450)$ are produced from the $\bar{D}^*\Sigma_c$ interaction, and one of them from $\bar{D}^*\Sigma^*_c$  interactions. Three of six bound states carry  opposite parities, which  are beyond the S-wave interactions. If only considered the states near the corresponding thresholds, the $P_c(4450)$ and the $P_c(4380)$ can be assigned as a $\bar{D}\Sigma_c^*$ and a $\bar{D}^*\Sigma_c$ molecular state, respectively.

Here, the cutoff is taken as free parameter for each interaction, even for each spin parity of an interaction. Physically, the cutoffs for different interaction can be different from each other while the same cutoff should be taken for certain interaction. Moreover, we should check if the bound states with oppositive parity, which correspond to P- or higher wave interactions, can be observed in the experimental observation channel $J/\psi p$. It is interesting to see that the cutoffs to produce the bound states with spin parities $3/2^-$ and $5/2^+$ from the  $\bar{D}^*\Sigma_c$ interaction are almost the same. In the next section, we will focus on $\bar{D}^*\Sigma_c$ interaction to study the possibility to interpret both LHCb pentaquarks as molecular states form this one interaction.

\section{The molecular states from $\bar{D}^*\Sigma_c$ interaction}\label{sec3}

In this section, we focus on the $\bar{D}^*\Sigma_c$ interaction, which threshold is close to the higher LHCb pentaquark $P_c(4450)$. Besides the $\bar{D}^*\Sigma_c$ channel, the  $J/\psi p$ channel where the LHCb pentaquarks were observed is also included to study the effect of the produced bound states on the $J/\psi p$ invariant mass spectrum. Since the $J/\psi p$ interaction is OZI suppressed, here we  assume the potential in this channel is zero. The transition between the $\bar{D}^*\Sigma_c$ and the $J/\psi p$ channel is described by  $D$ and $D^*$ exchanges~\cite{He:2016pfa}. By solving a coupled-channel Bethe-Salpeter equation, the results of the two-channel scattering in $3/2^-$ and $5/2^+$ waves are presented in Fig.~\ref{Fig:
DSigmaA}.
\begin{figure}[h!]
\begin{center}
\includegraphics[bb=112 50 303 300,clip, scale=0.8]{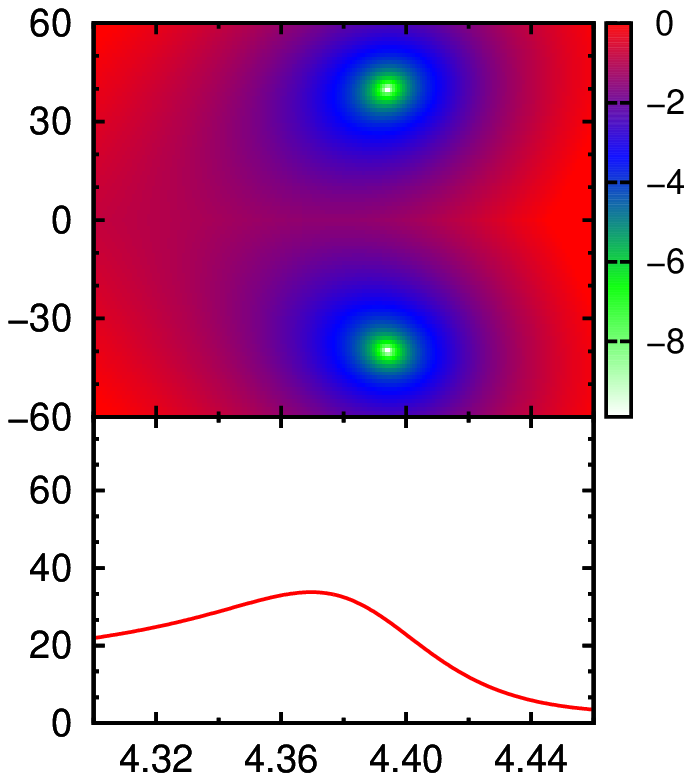}
\includegraphics[bb=157 50 330 300,clip, scale=0.8]{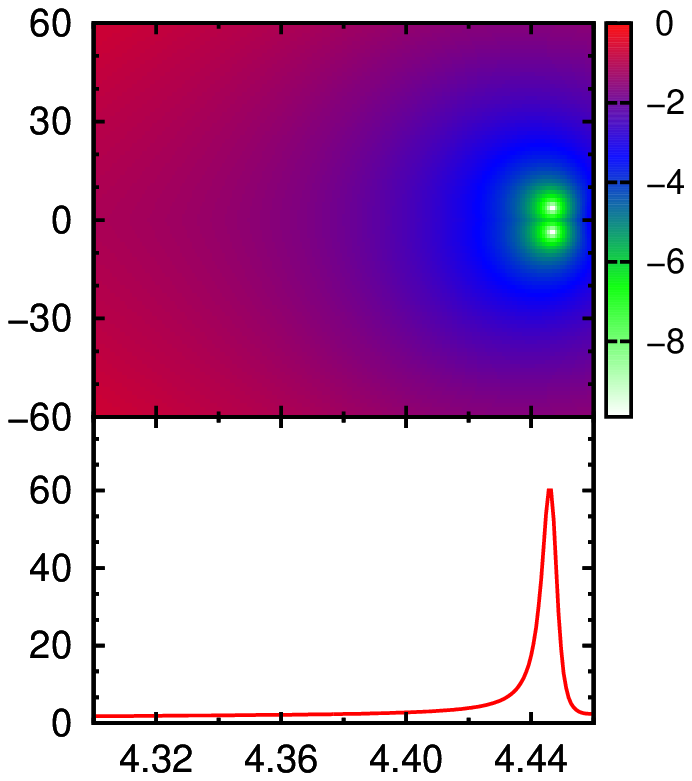}
\put(-160,5){\footnotesize \bf Re($z$) (GeV)}
\put(-293,55){ \rotatebox{90}{\footnotesize $\bm d\sigma/dW$}}
\put(-293,120){ \rotatebox{90}{\footnotesize \bf Im($z$) (MeV)}}
\put(-250,80){\color{blue}{\large$\bf\frac{3}{2}^-$}~(S and D waves)}
\put(-130,80){\color{blue}{\large$\bf\frac{5}{2}^+$}~(P and F waves)}
\caption{The $|1-V(z)G(z)|$ (upper panels) and the $J/\psi p$ mass spectrum (lower panels) for the $\bar{D}^*\Sigma_c$ interaction coupled with $J/\psi p$ channel at cutoff $\Lambda$=2 GeV. The results in $\frac{3}{2}^-$ wave (left panels) and $\frac{5}{2}^+$ wave (right panels) are drawn to the same scale.
\label{Fig: DSigmaA}}
\end{center}
\end{figure}

After the $J/\psi p$ channel is included, the bound state from the $\bar{D}^*\Sigma_c$ interaction in $5/2^+$ wave leaves the real axis and becomes a pole in the complex plane at $4446\pm 5i$ MeV.  Correspondingly, a narrow peak in the $J/\psi p$ mass spectrum is found near the $\bar{D}^*\Sigma_c$ threshold.  Since the opposite parity corresponds to the P and F waves, the  relatively weak interaction leads to small binding energy and width of this state. Its closeness to the $\bar{D}^*\Sigma_c$ threshold and narrowness suggest that it can be identified as the experimentally observed $P_c(4450)$.
In  $3/2^-$ wave, a pole at $4388+35i$ MeV is found, which peak is rather broad and far from the  $\bar{D}^*\Sigma_c$ threshold because of the relatively strong interaction in this partial wave.
The $5/2^+$-wave state is bound more loosely and narrower than the $3/2^-$-wave state, which is consistent with the experimental observations of the $P_c(4450)$ and the $P_c(4380)$~\cite{Aaij:2015tga}. We would like to note that though binding energy of $5/2^+$-wave state is much smaller than  that of  $3/2^-$-wave state, the heights of the former is even a little  larger than that of the latter. It makes the $5/2^+$-wave state is easy to  observe in experiment.

\section{Discussion and conclusion}\label{sec5}

In this work, the molecular states form the $\bar{D}\Sigma^*_c$, $\bar{D}^*\Sigma_c$ and $\bar{D}^*\Sigma^*_c$  interactions and their relation to  experimentally observed $P_c(4450)$ and $P_c(4380)$ are investigated in an quasipotential Bethe-Saltpeter approach.
We also make an explicit studies about the  $\bar{D}^*\Sigma_c$ interaction coupled with $J/\psi p$ channel.  Two poles around $4450$ and $4390$ MeV are produced  from the $\bar{D}^*\Sigma_c$ in $3/2^-$ and $5/2^+$ waves, respectively.  The peak for $5/2^+$ state has a comparable hight as that of $3/2^-$ state in the $J/\psi p$ invariant mass spectrum. Another bound state with spin-parity $J^P=3/2^-$ is produced from  the $\bar{D}\Sigma^*_c$ interaction. As suggested in Ref.~\cite{Chen:2016qju} existence of two or more resonant signals around 4380 MeV, especially those with spin parity $3/2^-$, can not be excluded due to the large width for the $P_c(4380)$ obtained here and in experiment.  Hence, the $P_c(4450)$ can be assigned as the molecular state with $5/2^+$ from the $\bar{D}^*\Sigma_c$ interaction while the $P_c(4380)$ may have more origins.

This project is partially supported by the Major State
Basic Research Development Program in China (No. 2014CB845405),
the National Natural Science
Foundation of China (Grants No. 11275235 and No.11675228).

\end{document}